\documentclass[aps,prc,twocolumn,nofootinbib,showpacs,preprintnumbers,amsmath,amssymb]{revtex4}

\usepackage{graphicx}
\usepackage{dcolumn}
\usepackage{bm}
\usepackage{epsfig}
\usepackage{longtable}

\def\fun#1#2{\lower3.6pt\vbox{\baselineskip0pt\lineskip.9pt
\ialign{$\mathsurround=0pt#1\hfil##\hfil$\crcr#2\crcr\sim\crcr}}}

\begin{document}

\preprint{}

\title{
Precise measurement of $\alpha_K$ and $\alpha_T$ for the 150.8-keV $E$3 transition in $^{111}$Cd: Test of internal-conversion theory
}

\author{N. Nica}
\email{nica@comp.tamu.edu}

\author{J.C. Hardy}
\email{hardy@comp.tamu.edu}

\author{V.E. Iacob}

\author{T.A. Werke}

\author{C.M. Folden III}

\author{L. Pineda}
\altaffiliation {REU summer student from New Mexico State University, Las Cruces, NM}

\affiliation{ Cyclotron Institute, Texas A\&M University, College Station, Texas 77843, USA}
\homepage{http://cyclotron.tamu.edu/}

\author{M.B. Trzhaskovskaya}
\affiliation{Petersburg Nuclear Physics Institute, Gatchina 188300, Russia}

\date{\today}

\begin{abstract}
We have measured the $K$-shell and total internal conversion coefficients, $\alpha_K$ and $\alpha_T$, for the 150.8-keV
$E$3 transition in $^{111}$Cd to be 1.449(18) and 2.217(26) respectively.  The $\alpha_K$ result agrees well with
Dirac-Fock calculations in which the effect of the $K$-shell atomic vacancy is accounted for; it extends our precision
tests of $\alpha_K$ calculations to $Z$ = 48, the lowest $Z$ yet measured.  However, the result for $\alpha_T$ disagrees
by about two standard deviations from the calculated $\alpha_T$ value, whether or not the atomic vacancy is included. 
 
\end{abstract}

\pacs{23.20.Nx, 27.60.+j}

\maketitle

\section{\label{sec:introd} INTRODUCTION}

Over the past decade we have published a sequence of papers \cite{Ni04,Ni05,Ni07,Ni08,Ni09,Ni14,Ha14}, in which we reported 
measurements of $K$-shell Internal Conversion Coefficients (ICCs) for $E$3 and $M$4 transitions in 6 nuclei with a precision
of $\pm$2\% or better. The motivation has been to test ICC theory, in particular its treatment of the $K$-shell vacancy left
behind by the emitted electron. What makes such precise measurements possible for us is our having an HPGe detector whose
relative efficiency is known to $\pm$0.15\% ($\pm$0.20\% absolute) over a wide range of energies: See, for example,
Ref.\,\cite{He03}.  By detecting both the $K$ x rays and the $\gamma$ rays from a transition of interest in the same
well-calibrated detector at the same time, we can avoid many sources of error.

By 2008, our early results from this program influenced a reevaluation of ICCs by Kib\'{e}di $et~al.$ \cite{Ki08}, who also
developed BrIcc, a new data-base obtained from the basic code by Band $et~al.$ \cite{Ba02} but, in conformity with our
conclusions, it employed a version of the code that incorporates the vacancy in the ``frozen orbital" approximation.  The
BrIcc data-base has been adopted by the National Nuclear Data Center (NNDC) and is available on-line for the determination
of ICCs.  Our experimental results obtained since 2008 continue to support that decision and have included transitions in
nuclei that cover the range $50<Z<78$.

We report here a measurement that extends the range down to $Z=48$.  We have measured the $\alpha_K$ and $\alpha_T$ values
for the 150.8-keV $E$3 transition in $^{111}$Cd to precisions of $\pm$1.7\% and $\pm$0.8\%, respectively. These
results are interesting not so much in distinguishing between models to account for the atomic vacancy -- the difference in
calculated ICCs between models that do and do not include the vacancy is rather small -- but in testing whether either
model comes close to describing the experimental data at all. Previous measurements of $\alpha_K$ and $\alpha_T$ for this transition
\cite{Lu60,Su85,Ne87} have been inconsistent with one another, scattering widely, but generally being significantly lower
than the corresponding calculated values.  Since the transition is hindered by $\sim$$10^4$ relative to the Weisskopf limit
\cite{Su85,Ba89}, the discrepancy has been associated with ICC anomalies observed for some other strongly hindered transitions.
However, we have previously demonstrated \cite{Ni07} that the $\alpha_K$ for the 127.5-keV $E$3 transition in $^{134}$Cs, a
similarly hindered case \cite{Ba89}, agrees well with calculations. Is the transition in $^{111}$Cd really discrepant and, if
so, does the disagreement with theory extend to both $\alpha_K$ and $\alpha_T$ values?

\section {\label{overview} Measurement Overview}

In our previous measurements we were dealing with decay schemes dominated by a single transition that can convert in the
atomic $K$ shell.  Under those conditions, if a spectrum of $K$ x rays and $\gamma$ rays is recorded, then the $K$-shell ICC for
the transition is given by
\begin{equation}
\alpha_K = \frac{N_K}{N_\gamma} \cdot \frac{\epsilon_\gamma}{\epsilon_K} \cdot \frac{1}{\omega_K},
\label{alpha1}
\end{equation}
where $\omega_K$ is the fluorescence yield; $N_K$ and $N_{\gamma}$ are the total numbers of observed
$K$ x rays and $\gamma$ rays, respectively; and $\epsilon_K$ and $\epsilon_\gamma$ are the
corresponding photopeak detection efficiencies.

The fluorescence yield for cadmium has been measured several times, with a weighted average quoted to $\pm$2.1\% \cite{Ka12}.
Furthermore, world data for fluorescence yields have also been evaluated systematically as a function of $Z$ \cite{Sc96}
for all elements with $10 \leq Z \leq 100$, and $\omega_K$ values have been recommended for each element in this range.  The
recommended value for cadmium, $Z$ = 48, is 0.842(4), which is consistent with the average measured value but has a smaller
relative uncertainty, $\pm$0.5\%.  We use this value.

\begin{figure}[t]
\epsfig{file=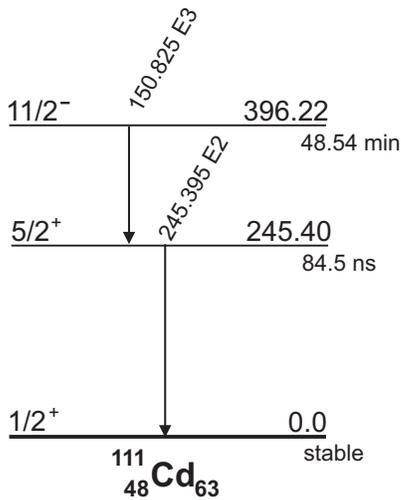,width=5.2cm}
\caption{Decay scheme for the 48.5-min isomer in $^{111}$Cd.  The data are taken from Ref.\,\cite{Bl09}.}
\label{fig1}
\end{figure}

The decay scheme of the 48.5-min isomer in $^{111}$Cd is shown in Fig.\,\ref{fig1}.  It does not have a single dominant
transition but rather a cascade of two, both of which can convert in the $K$ shell.  Thus, to extract the $\alpha_K$ value for
the 150.8 keV transition, we must use a modified version of Eq.\,\ref{alpha1}:
\begin{equation}
\alpha_{K150} = \frac{N_K}{N_{\gamma150}} \cdot \frac{\epsilon_{\gamma150}}{\epsilon_K} \cdot \frac{1}{\omega_K}-\alpha_{K245}
    \cdot \frac{N_{\gamma245}}{N_{\gamma150}} \cdot \frac{\epsilon_{\gamma150}}{\epsilon_{\gamma245}},
\label{alpha2}
\end{equation}
where the subscripts 150 and 245 on a quantity denote the transition -- either the 150.8-keV or 245.4-keV one -- to which
that quantity applies.  Note that the result we are seeking for $\alpha_{K150}$ now depends on $\alpha_{K245}$ but since the
245.4-keV transition is pure $E$2 and has higher energy, its $K$-conversion coefficient is about an order of magnitude smaller,
so the dependence of $\alpha_{K150}$ on $\alpha_{K245}$ is rather weak and does not seriously degrade the precision with which
the former can be determined.

Having a second transition in cascade with the 150.8-keV transition also offers an advantage: It allows the determination of
$\alpha_{T150}$ via the equation
\begin{equation}
(1+\alpha_{T150}) \cdot \frac{N_{\gamma150}}{\epsilon_{\gamma150}} = (1+\alpha_{T245}) \cdot \frac{N_{\gamma245}}{\epsilon_{\gamma245}}.
\label{alphatot}
\end{equation}
In this case, the fact that $\alpha_{T245}$ is much smaller than 1 ($\sim$0.06) and $\alpha_{T150}$ is greater than 1 ($\sim$2.2),
the experimental result obtained for $\alpha_{T150}$ is rather insensitive to the calculated value used for $\alpha_{T245}$.

Note that in Eqs.\,(\ref{alpha1}-\ref{alphatot}) all the $N$ values have to incorporate corrections for coincidence summing, which
for the two gamma rays must also include the effect of the angular correlation between them.  For $N_K$, the contributions from
impurities  must also be removed.

In our experiments, we detect the $\gamma$ rays and the $K$ x rays in the same HPGe detector, a detector whose efficiency has
been meticulously calibrated \cite{Ha02,He03,He04} to sub-percent precision, originally over an energy range from 50 to 3500 keV but
more recently extended \cite{Ni14} down to 22.6 keV, the average energy of silver $K$ x rays.  Over this whole energy region, precise
measured data were combined with Monte Carlo calculations from the CYLTRAN code \cite{Ha92} to yield a very precise and accurate detector
efficiency curve.  In our present study, the $\gamma$ rays of interest at 150.8 and 245.4 keV are well within the energy region for
which our efficiencies are known to a relative precision of $\pm$0.15\%.  The cadmium $K$ x rays lie between 23 and 27 keV, just
within our extended region of calibration so the detector efficiency for them can only be quoted to a relative precision of $\pm$1\%.

\section{\label{exp} Experiment}

We used the same experimental method and setup as in our previous measurements \cite{Ni04,Ni05,Ni07,Ni08,Ni09,Ni14,Ha14}.  Only those
details not covered in previous publications will be described here.

\begin{figure*}[t]
  \begin {minipage}[c]{0.67\textwidth}
\epsfig{file=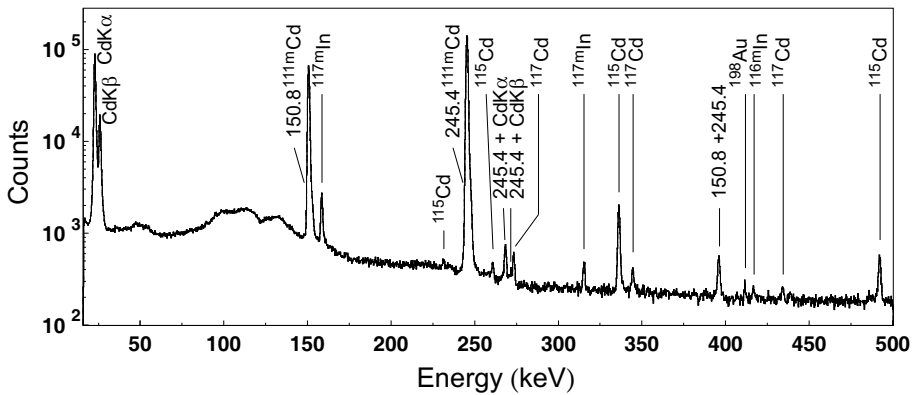,width=\textwidth}
\end{minipage}\hfill
\begin{minipage}[c]{0.3\textwidth}
\caption{Portion of the x- and $\gamma$-ray energy spectrum measured for 15 min with source S2.  Room background has been subtracted. Peaks are labeled by their $\beta$-decay parent;
the $\gamma$-ray peaks associated with $^{111m}$Cd decay also have their energies in keV marked.}
\label{fig2}
  \end{minipage}
\end{figure*} 

\subsection{\label{sprep} Source Preparation}

We produced $^{111}$Cd by neutron activation of a thin layer of 95.88\%-enriched $^{110}$CdO electroplated onto a 10-$\mu$m-thick foil
of 99.999\% pure aluminum.  The source material was obtained as cadmium metal powder from Trace Sciences International, which itemized its
chemical and isotopic impurities.  The former totaled less than 250 ppm and, of the latter, only $^{106}$Cd (0.01\%), $^{108}$Cd (0.02\%),
$^{114}$Cd (0.84\%) and $^{116}$Cd (0.16\%) are of potential relevance to our post-activation $\beta$-delayed $\gamma$-ray spectrum.

Two samples were prepared, each from $\sim$0.9 mg of the enriched metal foil.  Each foil was dissolved in 100 $\mu$L of 2-molar HNO$_3$ and evaporated
to dryness under argon gas to convert the metal into the nitrate form. The sample was then reconstituted with 5-10 $\mu$L of 0.1-molar HNO$_3$ and
$\sim$12 mL of anhydrous isopropanol.  We then transferred this solution to an electrodeposition cell, and electrochemically deposited the
$^{110}$Cd(NO$_3$)$_2$, using the molecular plating technique \cite{Pa62,Pa64}, onto a 10-$\mu$m-thick pure Al backing (99.999\% natural
aluminum; purchased from Goodfellow USA).  The deposition voltage was 350-700 V with a current density of $\sim$2 mA/cm$^2$.  Deposition times
ranged from 40 to 60 min.  After deposition, each target was rinsed in acetone for 5 min and baked in atmosphere at 200$^{\circ}$C for 30 min to
convert the $^{110}$Cd(NO$_3$)$_2$ to $^{110}$CdO.  The plating efficiencies were between 95 and 100\%.

We used identically made $^{\textup{nat}}$CdO targets to characterize the products instead of $^{110}$CdO because the analysis techniques led to
destruction of the targets.  Since the conditions for molecular plating were carefully controlled and are highly reproducible, we expect that
the chemical and physical properties of the natural and isotopically enriched targets were very similar.  The $^{\textup{nat}}$CdO targets were
analyzed with scanning-electron microscopy (SEM) and atomic-force microscopy (AFM) to examine the surface and ensure uniformity.  The average
variation in the thickness was $\sim$0.1\,$\mu$m as determined by AFM; and the images taken via SEM showed a mostly smooth, uniform layer of
$^{\textup{nat}}$CdO with few cracks or other irregularities.  Finally, an energy-dispersive x-ray-spectrometry analysis confirmed that the
$^{\textup{nat}}$CdO targets did indeed have a 1:1 cadmium-to-oxygen atomic ratio and that no major contaminants were introduced during the
molecular plating process.

One of the $^{110}$CdO targets (S1) had an areal density of 435(8) $\mu$g/cm$^2$; the other (S2) had 475(8) $\mu$g/cm$^2$.  Both were 1.7 cm
in diameter.  The sources were activated, one month apart, for 2\,h in a neutron flux of $\sim7\times10^{12}\,n$/(cm$^2$\,s) at the
TRIGA reactor in the Texas A\&M Nuclear Science Center.  After removal from the reactor each sample was conveyed to our measurement location,
where counting began approximately 4\,h after the end of activation. The initial activity from $^{111m}$Cd was determined to be $\sim$80\,kBq.

\subsection{\label{decaymeas} Radioactive decay measurements}

We acquired spectra with our precisely calibrated HPGe detector and with the same electronics used in its calibration \cite{He03}.  Our
analog-to-digital converter was an Ortec TRUMP$^{TM}$-8k/2k card controlled by MAESTRO$^{TM}$ software.  We acquired 8k-channel spectra at a
source-to-detector distance of 151~mm, the distance at which our calibration is well established.  Each spectrum covered the energy interval
10-2000 keV with a dispersion of about 0.25 keV/channel.

With source S1 we acquired 14 spectra over a two-week period; with S2 we acquired 10 spectra in one week.  Since $^{111m}$Cd has a half-life
of only 49 min, all but the first few spectra recorded in each case were used solely as an aid in identifying contaminant activities.  All
subsequent analysis was based on only two spectra -- the first 9-min spectrum taken from S1 and the first 15-min spectrum from S2 -- where the
$^{111m}$Cd activity was most dominant. It is the results from these two spectra that are reported here.

\section{\label{sec:analysis} Analysis}

\begin{table}[b]
\caption{\label{table1} The contributions of identified impurities to the energy region of the cadmium $K$ x-ray peaks.}
\vspace{2mm}
\begin{ruledtabular}
\begin{tabular}{llllll}
Source & Contaminant &  \multicolumn{4}{c}{Contribution to spectrum (\%)}  \\[1mm]
\cline{3-6}
& & &  \\[-2mm]
& & \hspace*{5mm}  & \multicolumn{1}{c}{S1} & \hspace*{5mm} & \multicolumn{1}{c}{\!\!\!S2}  \\
\hline  \\[-2mm]
$^{115}$Cd & In $K$ x rays & & 1.81(8) & & 2.71(6) \\
$^{117}$Cd & In $K$ x rays & & 0.113(12) & & 0.125(9) \\
$^{117}$In & Sn $K$ x rays & & 0.18(3) & & 0.21(3)  \\
$^{117m}$In & In+Sn $K$ x rays & & 0.603(9) & & 0.484(5)  \\
$^{116m}$In & In $K$ x rays & & 0.010(2) & & 0.006(1)  \\
\end{tabular}
\end{ruledtabular}
\end{table}

A portion of the S2 spectrum is presented in Fig.\,\ref{fig2}: It includes the x- and $\gamma$-ray peaks of interest from the decay of $^{111m}$Cd,
as well as a number of weak peaks from contaminant activities.  In our analysis of the data, we followed the same methodology as we did with previous
source measurements \cite{Ni04,Ni05,Ni07,Ni08,Ni09,Ni14,Ha14}.  We first extracted areas for essentially all the x- and $\gamma$-ray peaks in the
background-subtracted spectrum.  Our procedure was to determine the areas with GF2, the least-squares peak-fitting program in the RADWARE series
\cite{Rapc}.  In doing so, we used the same fitting procedures as were used in the original detector-efficiency calibration \cite{Ha02,He03,He04}.

Once the areas (and energies) of peaks had been established, we could identify all impurities in the $^{111m}$Cd spectrum and carefully check
to see if any were known to produce x or $\gamma$ rays that might interfere with the cadmium $K$ x rays or either of the two $\gamma$-ray peaks
of interest, at 150.8 and 245.4 keV.  As is evident from Fig.\,\ref{fig2}, even the weakest peaks were identified.  In all, we found 5 weak
activities that contribute to the cadmium x-ray region; these are listed in Table\,\ref{table1}.  No impurities interfere in any way with either
of the $\gamma$-ray peaks.

The count total recorded in the energy region of the cadmium $K_{\alpha}$ and $K_{\beta}$ x rays appears for each source in the top row of
Table\,\ref{table2}.  The impurity totals, derived from the percentage breakdowns listed in Table\,\ref{table1}, are given in the second
row of Table\,\ref{table2}.  Three more corrections need to be applied at this stage: The first relates to the shape of the x-ray peaks. As
explained in our previous papers (see, for example, Ref.\,\cite{Ni04}) we use a special modification of the GF2 program that allows us to sum
the total counts above background within selected energy limits.  To account for possible missed counts outside those limits, the program adds
an extrapolated Gaussian tail.  This extrapolated tail does not do full justice to x-ray peaks, whose Lorentzian shapes reflect the finite widths
of the atomic levels responsible for them.  To correct for this effect we compute simulated spectra using realistic Voigt functions to generate
the x-ray peaks, and we then analyze them with GF2, following exactly the same fitting procedure as is used for the real data, to ascertain how
much was missed by this approach.  The resultant correction factor appears as a percent in Table\,\ref{table2}. 

\begin{table}[t]
\caption{\label{table2}Corrections to the $^{111m}$Cd $K$ x rays and the 150.8- and 245.4-keV $\gamma$ rays, as well as the additional
information required to extract a value for $\alpha_{K150}$. }
\vspace{2mm}
\begin{ruledtabular}
\begin{tabular}{lll}
Quantity   &  \multicolumn{2}{c}{Value}  \\
\cline{2-3} \\[-2mm]
& \multicolumn{1}{c}{S1} & \multicolumn{1}{c}{S2} \\
\hline \\[-2mm]
\multicolumn{3}{l}{Cd ($K_{\alpha} + K_{\beta}$) x rays}  \\
~~Total counts & ~~\,$1.979(6)\times$$10^5$ & ~~\,$4.695(9)\times$$10^5$ \\
~~Impurities  & $-5.39(14)$$\times10^3$  & $-1.66(3)\times$$10^4$  \\
~~Lorentzian correction  &  +0.12(2)\%  &  +0.12(2)\% \\
~~Summing correction & +0.99(6)\%  & +0.99(6)\%  \\
~~Attenuation correction & +0.27(2)\% & +0.29(2)\%  \\
~~Corrected counts, $N_K$  & ~~\,$1.952(6)\times$$10^5$  & ~~\,$4.593(10)\times$$10^5$ \\
\hline \\[-2mm]
\multicolumn{3}{l}{$^{111}$Cd 150.8-keV $\gamma$ ray}  \\
~~Total counts & ~~\,$1.303(11)\times$$10^5$ & ~~\,$3.064(25)\times$$10^5$ \\
~~Summing correction &  +1.29(6)\%  &  +1.29(6)\% \\
~~Corrected counts, $N_{\gamma150}$ & ~~\,$1.320(12)\times$$10^5$ & ~~\,$3.104(25)\times$$10^5$ \\
\hline \\[-2mm]
\multicolumn{3}{l}{$^{111}$Cd 245.4-keV $\gamma$ ray}  \\
~~Total counts & ~~\,$3.024(22)\times$$10^5$ & ~~\,$7.082(45)\times$$10^5$ \\
~~Summing correction &  +0.86(3)\%  &  +0.86(3)\% \\
~~Corrected counts, $N_{\gamma245}$ & ~~\,$3.050(22)\times$$10^5$ & ~~\,$7.143(45)\times$$10^5$ \\
\hline \\[-2mm]
$N_K/N_{\gamma150}$ & ~~\,1.479(14) & ~~\,1.480(12) \\
$N_{\gamma245}/N_{\gamma150}$ & ~~\,2.311(27) & ~~\,2.301(24) \\
\end{tabular}
\end{ruledtabular}
\end{table}

The second correction takes account of true-coincidence summing.  Because the two transitions from $^{111m}$Cd occur in rapid succession, there is
a finite possibility that two $\gamma$ rays, or a $\gamma$ ray from one transition and an x ray from the other, can appear essentially simultaneously
in our detector and thus be recorded with an apparent energy that differs from 190.8 or 245.4 keV or from the $K$ x-ray energy.  If not corrected
for, these summing processes would falsely deplete the counts in the peaks of interest.  We can easily calculate the effects of this true-coincidence
summing (as distinct from random summing, which is negligible in our case) because the efficiency of our HPGe detector and the total-to-peak ratio
of its response function are both well known as a function of $\gamma$-ray energy \cite{He04}.  The resultant correction factor for the x ray appears
as a percent in the table.

The third correction arises from the finite thickness of our samples, which leads to some additional attenuation of the $K$ x rays relative to the
higher-energy $\gamma$ rays, over and above the calibrated relative efficiencies of the HPGe detector.  These percentage corrections are also listed
in the table, and the values for $N_K$, corrected for impurities, the Lorentzian shape, summing and relative attenuation, are given on the line below. 

The counts recorded in the two $\gamma$-ray peaks also appear in Table\,\ref{table2}.  They require no corrections for impurities or shape but they
are subject to summing corrections, so these are also given in the table.  In evaluating these corrections, we have taken account of the calculated angular
correlation between the two $\gamma$ rays, which favors small angles and thus serves to increase the summing probability slightly.  The corrected
values for $N_{\gamma 150}$ and $N_{\gamma245}$ are shown for both sources immediately below the summing correction.

In anticipation of their use in Eqs.\,(\ref{alpha2}) and (\ref{alphatot}), the ratios $N_K/N_{\gamma150}$ and $N_{\gamma245}/N_{\gamma150}$
also appear in Table\,\ref{table2}, where it can be seen that the results from both sources are entirely consistent with one another.  Consequently,
we use the weighted average of results from the two sources for all subsequent analysis: viz.
\begin{eqnarray}
N_K/N_{\gamma150} & = & 1.479(10)
\nonumber \\
N_{\gamma245}/N_{\gamma150} & = & 2.305(18)
\label{avrat}
\end{eqnarray}

     We deal next with the efficiency ratios, $\epsilon_{\gamma 150}/\epsilon_{\gamma 245}$ and $\epsilon_{\gamma 150}/\epsilon_K$, also required
by Eqs.\,(\ref{alpha2}) and (\ref{alphatot}).  The former can be obtained from our well-established detector efficiency curve obtained via CYLTRAN
Monte Carlo calculations \cite{He03}.  The latter requires a multi-step process. Following the same procedure as the one we used in analyzing the
decay of $^{119m}$Sn \cite{Ni14}, we employ as low-energy calibration the well-known decay of $^{109}$Cd, which emits 88-keV $\gamma$ rays and
silver $K$ x rays.  The latter are very close in energy to the cadmium $K$ x rays observed in the current measurement.  Note that we are not
distinguishing between $K_{\alpha}$ and $K_{\beta}$ x rays.  Scattering effects are quite pronounced at these energies and they are difficult to
account for with an HPGe detector when peaks are close together, so we have chosen as before to use only the sum of the $K_{\alpha}$ and $K_{\beta}$
x-ray peaks.  For calibration purposes, we consider each sum to be located at the intensity-weighted average energy of the component peaks---23.62
keV for cadmium and 22.57 keV for silver.  

    We obtain the required ratio, $\epsilon_{\gamma 150}/\epsilon_{K23.6}$ from the following relation:
\begin{equation}
\frac{\epsilon_{\gamma 150}}{\epsilon_{K23.6}} = \frac{\epsilon_{\gamma\,88.0}}{\epsilon_{K22.6}} \cdot
 \frac{\epsilon_{\gamma 150}}{\epsilon_{\gamma 88.0}} \cdot \frac{\epsilon_{K22.6}}{\epsilon_{K23.6}},
\label{effratio}
\end{equation} 
where here we have identified the $K$ x rays by their centroid energy for clarity.  We take the ratio $\epsilon_{\gamma 88}/\epsilon_{K22.6}$
= 1.069(8) from our previously reported measurement \cite{Ni14}.  The ratio $\epsilon_{\gamma 150}/\epsilon_{\gamma 88.0}$ = 0.8707(13) is
determined from our known detector efficiency curve calculated with the CYLTRAN code \cite{He03}, while $\epsilon_{K22.6}/\epsilon_{K23.6}$ =
0.9849(9) comes from a CYLTRAN calculation as well but in an energy region with higher uncertainty.  Nevertheless, the energy span is so small that
the uncertainty on the ratio is also very small ($\sim$0.1\%).

The two efficiency ratios required by Eqs.\,(\ref{alpha2}) and (\ref{alphatot}) are thus:
\begin{eqnarray}
\frac{\epsilon_{\gamma 150}}{\epsilon_{\gamma 245}} & = & 1.3123(14)
\nonumber \\
\frac{\epsilon_{\gamma 150}}{\epsilon_K} & = & 0.917(7).
\label{effrat}
\end{eqnarray}

\section{\label{sec:results} Results and Discussion}

The $K$-shell ICCs for the two transitions from $^{111m}$Cd can be related to one another if we substitute the measured ratios from
Eqs.\,(\ref{avrat}) and (\ref{effrat}) into Eq.\,(\ref{alpha2}); the result is
\begin{equation}
\alpha_{K150} = 1.610(18) - 3.025(24)~\alpha_{K245}.
\label{alpharel}
\end{equation}
Because $\alpha_{K245}$ is about an order of magnitude smaller than $\alpha_{K150}$, its contribution to this equation is minor.  Nonetheless its value
must be calculated in order to obtain a value for $\alpha_{K150}$.  Our ICC calculations are made within the Dirac-Fock framework with the option either
to ignore the $K$-shell vacancy or to include it in the ``frozen orbital" approximation \cite{Ba02}.  For $\alpha_{K245}$ the two different calculations
yield values that differ by less than 1\%: 0.05309 (no vacancy) and 0.05343 (vacancy included).  So as not to prejudice our result for the 150.8-keV
transition, we adopt the value 0.05326(17), which encompasses both possibilities.  Substituting this value into Eq.\,(\ref{alpharel}) we obtain the
result:
\begin{equation}
\alpha_{K150} = 1.449(18),
\label{alphaK}
\end{equation}
where the uncertainty in $\alpha_{K245}$ makes a negligible contribution to the $\alpha_{K150}$ uncertainty.

Next, the total ICCs for the two transitions can be related to one another by substitution of Eqs.\,(\ref{avrat}) and (\ref{effrat}) into
Eq.\,(\ref{alphatot}); in this case the result is
\begin{equation}
\alpha_{T150} = 3.025(24)~(1+\alpha_{T245}) - 1.
\label{alphatotrel}
\end{equation}
The total ICC for the 245.4-keV transition is calculated to be 0.06333 if the vacancy is ignored and 0.06368 with the vacancy included, so again we
choose the average with an assigned uncertainty that encompasses both values: 0.06351(17).  Substituting this value into Eq.\,(\ref{alphatotrel}) we
obtain:
\begin{equation}
\alpha_{T150} = 2.217(26).
\label{alphatotfin}
\end{equation}
As with the $K$-shell ICCs, the uncertainty in $\alpha_{T245}$ has no impact on the uncertainty attached to $\alpha_{T150}$.

There has been only one previous measurement of $\alpha_{K150}$, by N\'{e}meth and Veres \cite{Ne87}, who also used an HPGe detector to record
both the x and $\gamma$ rays.  Their result, 1.29(11), is statistically consistent with ours but is a factor of 6 less precise -- see
Eq.\,(\ref{alphaK}).

The situation is more complicated for $\alpha_{T150}$, of which there have been three previous measurements.  The earliest, by Lu \cite{Lu60}, is from
1960: It was performed with a source placed in the well of a large NaI(Tl) scintillator, and produced a result, averaged over two independent runs,
of 2.286(9), with an astonishing $\pm$0.4\% uncertainty.  The two more-recent measurements, which both appeared in the 1980s, employed HPGe detectors
to obtain the values 1.76(8) \cite{Su85} and 1.98(5) \cite{Ne87}.  All three of these references were included in the 2002 survey of internal
conversion coefficients published by Raman {\it et al.}\cite{Ra02}, but the Lu result appeared there as 2.29(3), the survey authors having decided to increase
its uncertainty.  Finally Raman {\it et al.}~``adopt" 2.12(11) as the recommended value for $\alpha_{T150}$.  Strikingly, none of the three previous
measurements agree with one another or with our new result, which appears in Eq.\,(\ref{alphatotfin}); however, our result does agree with the
value adopted in the 2002 survey \cite{Ra02} and has an uncertainty smaller by a factor of 4.

\begin{table}[b]
\caption{\label{table3}Comparison of the measured $\alpha_K$ and $\alpha_T$ values for the 150.853(15)-keV $E$3 transition from $^{111m}$Cd with
calculated values based on two different theoretical models, one that ignores the $K$-shell vacancy and one that deals with it in ``frozen orbital" (FO)
approximation \cite{Ba02}. Shown also are the percentage deviations, $\Delta$, from the experimental value calculated as (experiment-theory)/theory.  For a
description of the various models used to determine the conversion coefficients, see Ref.\,\cite{Ni04}.}
\vspace{2mm}
\begin{ruledtabular}
\begin{tabular}{lllll}
\multicolumn{1}{l}{Model}  & \multicolumn{1}{c}{~~$\alpha_K$} & \multicolumn{1}{c}{~~$\Delta$(\%)} & \multicolumn{1}{c}{~~$\alpha_T$} & \multicolumn{1}{c}{~~$\Delta$(\%)} \\
\hline \\[-3mm]
Experiment & 1.449(18)  & & 2.217(26) &  \\
Theory: & & \\
~~~No vacancy  & 1.425(1) & +1.7(12) & 2.257(1) & $-1.8(12)$ \\
~~~Vacancy, FO  & 1.451(1) & $-0.1(12)$ & 2.284(1) & $-2.9(12)$ \\
\vspace{-10.pt}
\end{tabular}
\end{ruledtabular}
\end{table}

In Table\,\ref{table3}, our measured $\alpha_K$ and $\alpha_T$ results for the 150.8-keV $E$3 transition are each compared with two theoretical
values, one that was calculated without accounting for the atomic vacancy and one that included the vacancy in the ``frozen orbital" (FO) approximation \cite{Ba02}.  The
percentage deviations given in the table show excellent agreement between our $\alpha_K$ measurement and the calculation that includes provision for the atomic
vacancy.  This outcome is consistent with our previous five precise $\alpha_K$ measurements on $E$3 and $M$4 transitions in $^{119}$Sn \cite{Ni14,Ha14}, $^{134}$Cs
\cite{Ni07,Ni08}, $^{137}$Ba \cite{Ni07,Ni08}, $^{193}$Ir \cite{Ni04,Ni05} and $^{197}$Pt \cite{Ni09}, all of which agreed well with the FO calculations, and
disagreed -- some by many standard deviations -- with the no-vacancy calculations.

Our $\alpha_T$ result does not lead to such a simple conclusion: It is lower than both calculations, with the worst disagreement ($\sim$2.5$\sigma$) being with
the FO calculation. One possible explanation arises from the fact that the 150.8-keV transition is hindered by a factor of $10^4$ relative to the
single-particle Weisskopf estimate \cite{Ba89}.  Under such conditions, one could expect to encounter ``penetration", which is a dynamic effect associated with
the change from transition electromagnetic potentials used for a point nucleus to transition potentials required for a realistic finite-sized nucleus \cite{Li86}.
For unhindered electric transitions, the penetration effect is not significant, but it may reach several percent for magnetic transitions.  The effect is included
in our ICC calculations by an approximation based on the surface-current model \cite{Ra02} but it is done uniformly with all nuclei and all transitions.  For strongly
hindered transitions, the penetration effect can become more important, giving rise to non-negligible nuclear matrix elements in the expressions for the ICCs.  In
this way these particular ICCs become dependent on nuclear structure details and nuclear transition dynamics. 

Calculations of penetration factors for specific nuclei were attempted decades ago, but mainly for $M$1 and $E$1 transitions (see for example Refs.\,\cite{Li86,Kr62,
Le65, Kr74}).  To our knowledge, there are no convincing theoretical results for $E$3 transitions. Nevertheless, we may speculate that the discrepancy we observe
between experiment and theory for $\alpha_T$ is likely due to penetration.  Is it reasonable that the effect is evident for $\alpha_T$ and not for $\alpha_K$?
Yes, Krpi\'{c} and Ani\u{c}in \cite{Kr74} argue (for $M$1 transitions) that under certain conditions the penetration can be ``hidden"
by a cancellation that only affects one atomic shell and not the others.  It is plausible that this can explain our results; certainly no more definitive
explanation is possible.

\section{Conclusions}
Our measurement of the $K$-shell internal conversion coefficient for the 150.8-keV $E$3 transition from $^{111m}$Cd has yielded a value, $\alpha_K$ = 1.449(18), that
agrees with a version of the Dirac-Fock theory that includes the atomic vacancy.  It disagrees (by $\sim$1.5$\sigma$) with theory if the vacancy is ignored.  
This result is consistent with the conclusion reached from our previous five precise ICC measurements, and extends the validity of that conclusion down to $Z$
= 48. 

Our result for the total ICC of this transition, $\alpha_T$~=~2.217(26), is lower than theory by several percent.  Although we cannot draw any definitive conclusion
as to the cause of this discrepancy, we suggest that it may be due to penetration effects that have become noticeable because the transition is particularly
hindered.

\vspace{7mm}

\begin{acknowledgments}

We wish to thank the staff of the Texas A\&M Nuclear Science Center for their help in conducting neutron
activations.  This material is based upon work supported by the U.S. Department of Energy, Office of Science,
Office of Nuclear Physics, under Award Number DE-FG03-93ER40773, and by the Robert A. Welch Foundation under
Grant No.\,A-1397.

\end{acknowledgments}

\end{document}